\documentclass[aps,prd,floatfix,12pt,tightenlines,preprintnumbers,nofootinbib]{revtex4}

\catcode`@=11
\renewcommand\NAT@sort@cites[1]{\edef\NAT@cite@list{#1}}

\def\l@@sections#1#2#3#4{%
 \begingroup
  \everypar{}%
  \@nameuse{l@f@#2}%
  \set@tocdim@pagenum{#4}%
  \global\@tempdima\csname tocdim@#2\endcsname
  \leftskip\csname tocleft@#2\endcsname\relax
  \dimen@\csname tocleft@#1\endcsname\relax
  \parindent-\leftskip\advance\parindent\dimen@
  \rightskip\tocleft@pagenum plus 1fil\relax
  \skip@\parfillskip\parfillskip\z@
  \let\numberline\numberline@@sections
  \ignorespaces#3\unskip\nobreak\hskip\skip@
  \hb@xt@\rightskip{\hss\unhbox\@tempboxa\hskip 5pt}\hskip-\rightskip\hskip\z@skip
  \par
  \expandafter\aftergroup\csname tocdim@#2\endcsname\expandafter
 \endgroup\the\@tempdima\relax
}%

\def\l@f@section{%
 \addpenalty{\@secpenalty}%
 \tightenlines@sw{%
  \addvspace{0.3em plus\p@}%
 }{%
  \addvspace{1.0em plus\p@}%
 }%
 \bf
}%

\def\make@footnotetext#1{%
  \reset@font\footnotesize
  \interlinepenalty\interfootnotelinepenalty
  \splittopskip\footnotesep
  \splitmaxdepth\dp\strutbox
  \set@footnotewidth
  \@parboxrestore
  \protected@edef\@currentlabel{%
   \csname p@footnote\endcsname\@thefnmark
  }%
  \color@begingroup
   \@makefntext{%
    \baselineskip=12pt
    \parskip=12pt
    \setbox \strutbox \hbox{\vrule height 10pt depth 12pt width 0pt}
    \rule\z@\footnotesep\footnotesep=12pt \ignorespaces#1\@finalstrut\strutbox
    \@finalstrut\strutbox
   }%
  \color@endgroup
 \minipagefootnote@drop
}%

\def\@email#1#2{%
 \endgroup
 \vskip 12pt
 \@AF@join{#1\href{mailto:#2}{#2}}%
}%

\catcode`@=12

\newcommand{\figurelabel}[1]{\label{#1}}
\newcommand{\figuremark}[1]{Figure~\ref{#1}}

\newcommand{\be}{\begin{equation}}
\newcommand{\ee}{\end{equation}}
\newcommand{\bue}{\begin{displaymath}}
\newcommand{\eue}{\end{displaymath}}

\input epsf.sty

\begin{document}
\title{Inflation and cosmological perturbations\footnote{To
    be published in {\it The Future of Theoretical Physics and
    Cosmology} (eds: G.W. Gibbons, E.P.S. Shellard, and S.J. 
    Rankin), Proceedings of the Stephen Hawking 60th Birthday
    Conference, Cambridge, UK, 7-11 January 2002.}}
\preprint{MIT-CTP-3340}
\author{Alan H. Guth}
\email{guth@ctp.mit.edu}
\affiliation{Center for Theoretical
Physics,  Laboratory for Nuclear Science and Department of
Physics,\\
Massachusetts  Institute of Technology, Cambridge,
Massachusetts 02139}
\date{August 5, 2002}

\begin{abstract}
This talk, which was presented at Stephen Hawking's 60th birthday
conference, begins with a discussion of the early development of
the theory of inflationary density perturbations.  Stephen played
a crucial role in this work, at every level.  Much of the
foundation for this work was laid by Stephen's 1966 paper on
cosmological density perturbations, and by his 1977 paper with
Gary Gibbons on quantum field theory in de Sitter space.  Stephen
was a major participant in the new work, and he was also
a co-organizer of the 1982 Nuffield Workshop, where divergent
ideas about inflationary density perturbations were thrashed
about until a consensus emerged.  In the second part of the talk I
summarize the recent observational successes of these
predictions, I present a graph of the probability distribution
for the time of last scattering for CMB photons, and I summarize
a recent theorem by Borde, Vilenkin, and me which shows that
although inflation is generically eternal to the future,
an inflationary region of spacetime must be incomplete in null
and timelike past directions.

\end{abstract}

\maketitle

\tableofcontents

\section{The origin of inflationary fluctuations}

Since the topic of inflation and density perturbations pretty
much came to fruition at the Nuffield Workshop \cite{Nuffield}
almost 20 years ago, and since I was one of the attendees, I
thought I would spend about half my talk reminiscing about the
history of those events.  The three-week period of the Nuffield
meeting was certainly among the most exciting times of my life.  
I will always cherish the memories of those events, and I will
always be grateful to Stephen for the important role that he
played in making them happen.

When ``old inflation'' \cite{old-inflation} was first invented,
we knew at once that we could identify at least one possible
source of density perturbations: namely, the random nucleation of
bubbles.  In that model the phase transition was assumed to be
strongly first order, so it ended with bubbles of the new phase
nucleating and colliding, producing some kind of inhomogeneous
froth similar to boiling water.  But it was of course not clear
immediately whether these density perturbations would have the
right properties to serve as the seeds for structure formation. 
Stephen played an important role in sorting out this question,
as one of the key papers was that of Hawking, Stewart, and
Moss \cite{HMS}.  Once the issue was understood, the answer was
clear.  The density perturbations created in the old inflationary
model were ridiculously large --- colossal --- so there was
absolutely no hope of getting a universe from it that looked like
ours. 

Fortunately, inflation was resurrected by the invention of ``new
inflation'' by Linde \cite{Linde-new}, and Albrecht and Steinhardt
\cite{Albrecht-Steinhardt}, and it was in that context that the
question of density perturbations could first be intelligently
addressed.  For new inflation, however, the question of density
perturbations was much more subtle. 

The first I remember hearing about this was at a meeting at
Moriond in March of 1982. I chatted with Michael Turner, and I
guess he had been in touch with Jim Bardeen.  As far as I know,
Jim was one of first people to really worry about this issue. 
Jim had realized that in the context of the new inflationary
model, any density perturbations present before inflation would
be suppressed by an enormous exponential factor as they evolved
through the inflationary regime.  This meant that if the model
was to have any chance of viability, some mechanism would have to
be found to allow the density perturbations to survive. 
According to Michael, Jim was exploring the possibility that
there might be some mechanism at the very end of inflation that
would amplify the density perturbations which had been suppressed
out of sight by the process of inflation itself.  At about the
same time I had a conversation with Dave Schramm, who was also
aware of this problem, and who thought that maybe we would have
to learn about turbulence to understand how density perturbations
could survive inflation.

In May 1982, there was a chance meeting in Chicago of three
physicists interested in these questions: Michael Turner, Paul
Steinhardt, and Stephen Hawking.  Michael was of course based in
Chicago, as he still is, Paul was visiting to give a colloquium,
and Stephen was by chance visiting at the same time to
collaborate with Jim Hartle.  It had been five years since
Stephen had written his famous paper \cite{Gibbons-Hawking} with
Gary Gibbons on quantum field theory in de Sitter space, so
Stephen was well-versed in the role that quantum fluctuations can
play in an exponentially expanding space.  In that paper Gary and
Stephen had shown the now-classic result that quantum
fluctuations in de Sitter space produce thermal radiation with a
temperature
\be
T_{\rm GH} = {H \over 2 \pi} \ ,
\ee
where $H$ is the Hubble constant of the de Sitter space, and I
use units for which $\hbar = c = k_{\rm Boltzmann} = 1$.  Stephen
was of course also one of the world's experts on cosmological
density fluctuations, going back to his seminal work on the
evolution of density perturbations in 1966 \cite{Hawking-1966}. 
I was of course not part of this meeting, but I learned about it
from Michael and Paul while I was gathering material for my
popular level book\footnote{In the book I erroneously stated that
the meeting took place in April.  This was based on some early
discussions, but later Paul checked his travel records and
discovered that his trip extended from May 5 to 7, 1982.}
\cite{mybook}, of which the Nuffield Workshop was one of the
highlights.  Apparently not much in the way of details was
discussed at this meeting, but the three physicists discussed a
crucially important new idea: the possibility that the structure
of the cosmos originated in quantum fluctuations.  From the point
of view of unity in physics, this is a truly breathtaking idea,
proposing that the same quantum phenomena that are central to the
study of atoms and subatomic physics are also responsible for the
largest structures known to humans.  The meeting reportedly ended
when Stephen left to shop at F.A.O. Schwarz, the famous toy
store.\footnote{While the idea that the large-scale scale
structure of the universe originated from quantum fluctuations
was, so far as I know, new in the West, the idea had been
explored in the Soviet Union more than a year earlier, when
Mukhanov and Chibisov \cite{Mukhanov-Chibosov} studied quantum
fluctuations in the Starobinsky \cite{Starobinsky-inflate1,
Starobinsky-inflate2} version of inflation.}

\tracingoutput=0

Later in May, I learned from Paul Steinhardt that he and Michael
had come up with a plausible way of estimating the effects of
quantum fluctuations, in the context of what was then the
universally favored grand unified theory, the minimal SU(5)
model.  They concluded that the spectrum would be
scale-invariant, which is what cosmologists favored
\cite{Harrison,Zeldovich}, but that $\delta \rho / \rho$, the
fractional perturbations in the mass density, would be about
$10^{-16}$.  The answer that we all wanted was about $10^{-4}$, so
they knew they were not there yet, but they were still trying. 

On June 7, 1982, Stephen gave a seminar in Princeton about his
density perturbation calculations. I was not there either, but
Paul Steinhardt was.  The day after the talk, Paul and I had a
long telephone conversation in which he gave me a detailed
summary of what Stephen had said.  I took notes while on the
phone, and I still have them.  In contrast to Paul's and
Michael's result of $10^{-16}$, Stephen's calculations gave the
result of $\delta \rho/\rho \sim 10^{-4}$, exactly what was
wanted.  Shortly afterward Stephen circulated a preprint
\cite{Hawking-preprint} summarizing these calculations.  I
learned from Stephen that his preprint was actually written
before his trip to the U.S. at the beginning of May, although its
date is listed as June, when the typing was completed.  The
abstract of that preprint explained that irregularities would be
produced in inflationary models by quantum fluctuations in the
scalar field as it ran down the hill of an effective potential
diagram.  The abstract continued:
\begin{quote} 
These would lead to fluctuations in the rate of expansion which
would have the right spectrum and amplitude to account for the
existence of galaxies and for the isotropy of the microwave
background. 
\end{quote} 
So it was a big success!  Inflation actually worked, accounting
not only for the large-scale uniformity of the Universe, but also
for the spectrum of density fluctuations needed to explain the
tapestry of cosmic structure.

When Paul and I were discussing Stephen's Princeton seminar on
the telephone, Paul relayed to me essentially all the equations
that Stephen had presented.  These began with equations relating
the perturbations during inflation to the quantum fluctuations of
the scalar field, as described by the scalar field two-point
function.  We did not quite understand what Stephen was talking
about, but we assumed he was probably right.  Stephen also
described --- in a couple of lines --- how density perturbations
would evolve from the beginning of inflation up to the present
day.  These were the sort of calculations that Stephen had done
long before; he knew the answers, so he just wrote them down. 
Again Paul and I assumed that Stephen must have known what he was
talking about.  There was one step in the calculation, however,
where Stephen calculated the time derivative of the scalar field,
as it rolled down the hill of the potential energy diagram.  That
was really the only step in the calculation that I was capable at
the time of understanding, but as far as Paul and I could tell,
Stephen got it wrong.  If we did this calculation our way, the
answer became larger by about a factor of $10^4$, giving $\delta
\rho/\rho$ about equal to one.  So Paul and I were very skeptical
of Stephen's result, although, since we did not understand the
rest of the calculation, we were not really sure what Stephen was
thinking, or whether we were correctly interpreting his
equations.  From our point of view Stephen's result looked wrong,
but we were not absolutely sure.

I then began working on this in detail with So-Young Pi.  We
closely followed the pattern of Stephen's calculation, trying to
make sure that we understood each step.  A key element in
Stephen's approach was to think of the primary driving force
behind the density fluctuations as the fluctuation $\delta t
(\vec x)$ in the time at which inflation ends in different
places.  The description of density perturbations depends very
much on the coordinate system (i.e., the choice of gauge) in
which they are described, so the papers on this subject that came
out of the Nuffield meeting \cite{Hawking-pub, Starobinsky,
Guth-Pi, BST} do not all look the same.  While all of these
papers got the same answer, the calculations are difficult to
compare.  The papers by Stephen, Starobinsky, and the one by
So-Young and me all focussed on the time delay function $\delta t
(\vec x)$, but they still used somewhat different ways of
connecting this function to the final result.  The paper by
Bardeen, Steinhardt, and Turner used a more complete integration
of the full set of equations from perturbative general
relativity.  I have always found that the time-delay approach,
which we learned from Stephen, is the clearest way to understand
density fluctuations, as long as one is interested in models with
a single scalar field that rolls slowly, so that the Hubble
parameter $H$ can be treated as a constant during the period when
the perturbations of the relevant wavelengths are generated.  The
models that interested us in 1982 fell within these restrictions,
although the generality of the Bardeen-Steinhardt-Turner approach
(and further elaborations --- see for example \cite{MFB}) is
needed to analyze the wider range of models that have since
become relevant.

By the way, Stephen's paper on density perturbations
\cite{Hawking-pub} turns out to be his third most cited
paper of all time, so I am personally rather proud to have been
vaguely involved in Stephen's third most cited paper! I made a
list of Stephen's most cited papers from yesterday's listings on
SPIRES (http://www.slac.stanford.edu/spires/hep/), and I was
flabbergasted to see that Stephen has 33 papers that have more
than a hundred cites; I think this is really astounding.  Cites
do not mean everything, of course, but when you see this many,
you can be sure they mean something!

\section{The 1982 Nuffield workshop}

\begin{figure}[p]
\epsfbox{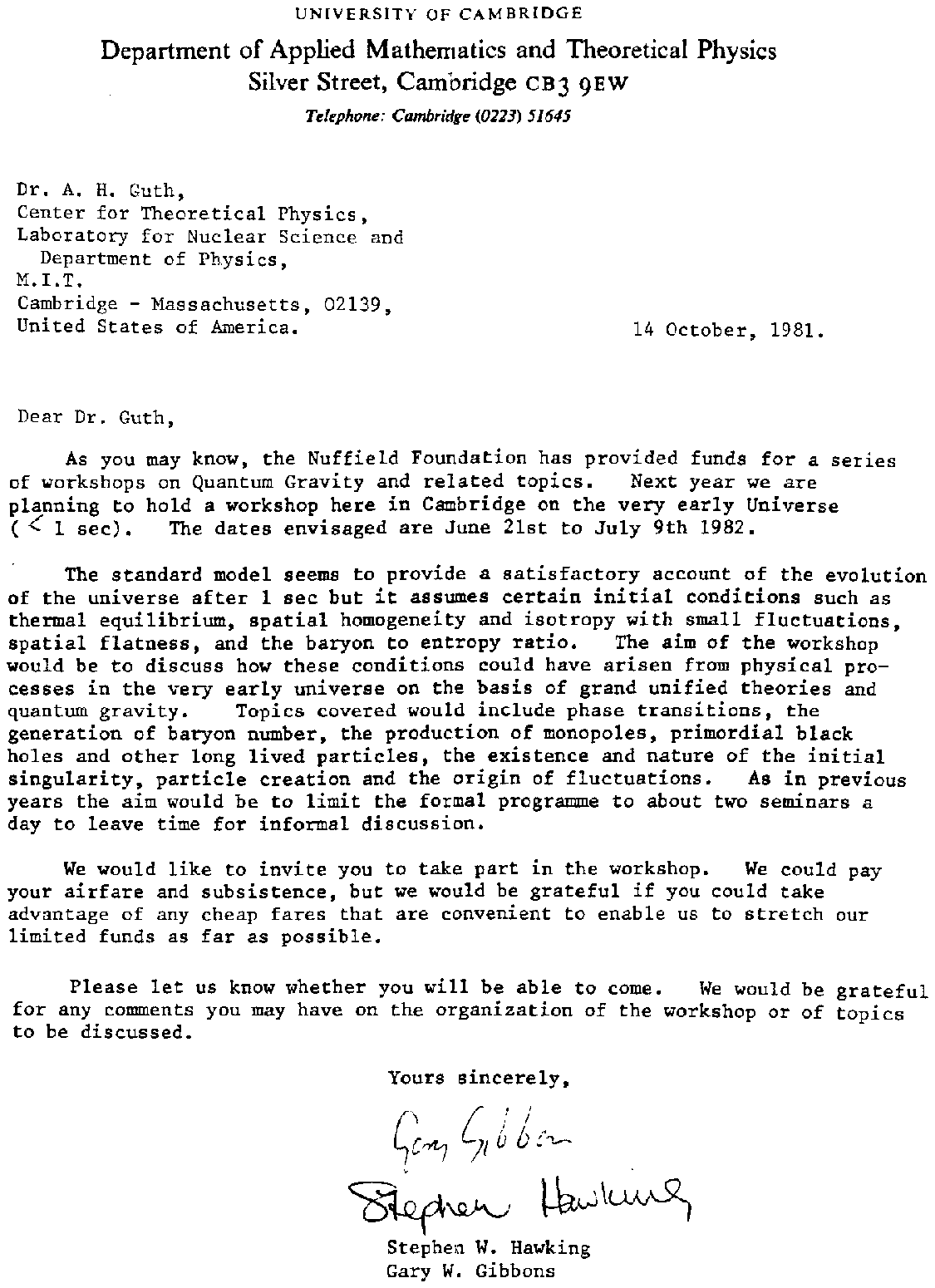}
\caption{Invitation to the Nuffield workshop on the Very Early Universe.}
\figurelabel{Nuffield}
\end{figure}

Chronologically, the next event --- the big event --- was the
Nuffield Workshop, which began on Monday, June 21, 1982.  I still
have my original invitation from the organizers of the meeting,
Gary Gibbons and Stephen (see \figuremark{Nuffield}).  I, of
course, didn't care about Gary, but I was so impressed that I
received a letter that was actually signed by the famous Stephen
Hawking (or at least by someone authorized to use his signature)
that I saved it for 20 odd years.  The key paragraph in the
letter stated the premise of the meeting, which with historical
hindsight seems very impressive.  It is a really good description
of where cosmology was at then, and some of these problems are
still problems that we are talking about now: 
\begin{quote} 
The standard model seems to provide a satisfactory account of the
evolution of the universe after 1 sec but it assumes certain
initial conditions such as thermal equilibrium, spatial
homogeneity and isotropy with small fluctuations, spatial
flatness, and the baryon to entropy ratio. The aim of the
workshop would be to discuss how these conditions could have
arisen from physical processes in the very early universe on the
basis of grand unified theories and quantum gravity.  Topics
covered would include phase transitions, the generation of baryon
number, the production of monopoles, primordial black holes and
other long lived particles, the existence and nature of the
initial singularity, particle creation and the origin of
fluctuations. 
\end{quote} 
What I want to talk about today is the success of this conference
in making real progress on the issue of the primordial density
perturbations.

I still have the transparencies from the talk I gave at the very
end of the conference, which summarizes what I knew then about
density fluctuations.  \figuremark{opening} shows my opening
transparency from that talk. I have to confess that the very
first thing on the transparency is the potential energy diagram
of the minimal SU(5) grand unified theory with a Coleman-Weinberg
potential, which at that time we all knew was the correct theory. 
Its amazing that after 21 years of research we now know so much
less --- but I guess that is a form of progress, too.

\begin{figure}[ht]
\centerline{\epsfbox{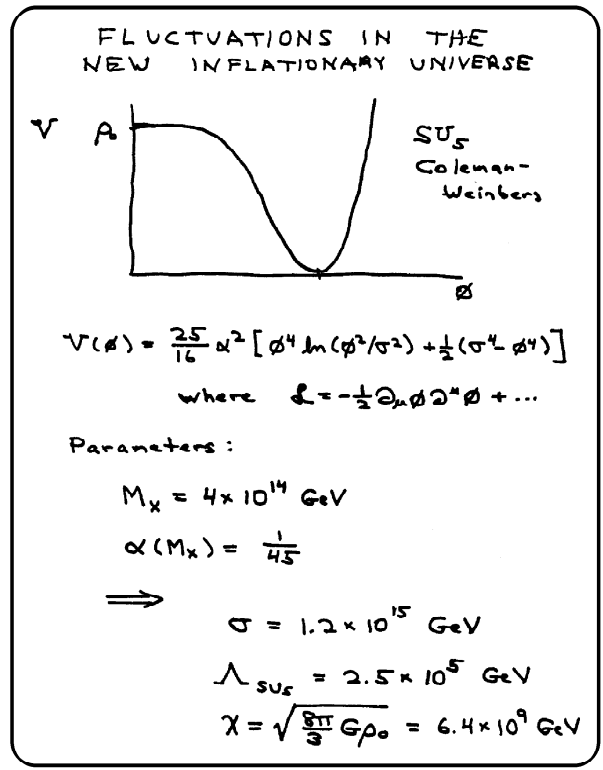}}
\caption{The opening transparency from my talk at the Nuffield
Workshop on {\it Fluctuations in the New Inflationary Universe}.}
\figurelabel{opening}
\end{figure}

In that same talk, I tried to describe the flow of thoughts on
the subject, so I made a transparency to summarize the evolution
of our thinking on the values of $\delta \rho / \rho$ as a
function of time in 1982 (see \figuremark{deltarho}).  We were
evaluating $\delta \rho/\rho$ for each mode at ``second
Hubble crossing,'' labeled on the chart as $2t = \ell$.  Here
$\ell$ is the wavelength in physical (as opposed to comoving) units,
and $2t$ in the radiation-dominated era is equal to $H^{-1}$, the
Hubble length.  (This is the second Hubble crossing, because for
each mode there was also a time during inflation when the
physical wavelength, which grows exponentially during inflation,
crosses the Hubble length, which is approximately constant during
inflation.)  The scale-invariant Harrison-Zeldovich spectrum 
is the spectrum for which $\delta \rho/\rho$ at second Hubble
crossing has the same value of each mode, and although we were
disagreeing on the amplitude, we were all getting approximately
scale-invariant spectra.

\begin{figure}[ht]
\centerline{\epsfbox{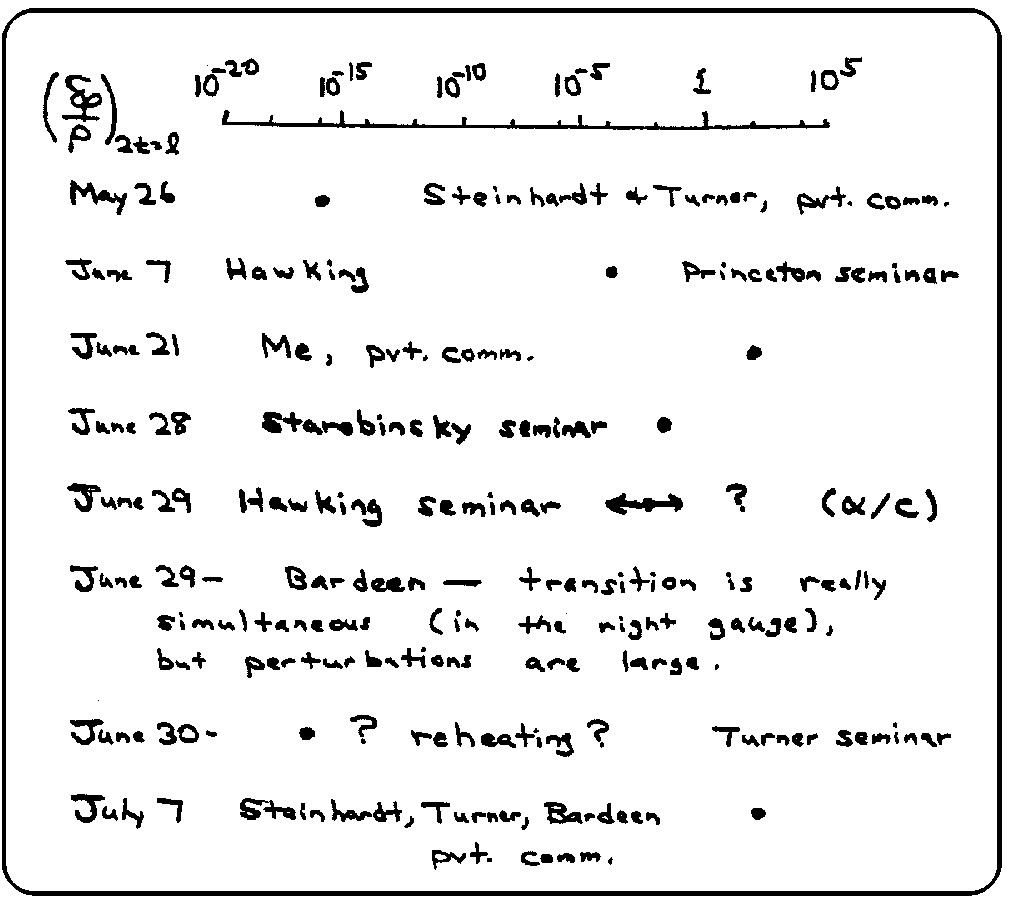}}
\caption{From my Nuffield Workshop talk: evolution of theoretical
predictions for the amplitude of the density fluctuations from
inflation.}
\figurelabel{deltarho}
\end{figure}

The chart starts at May 26, 1982, when Paul Steinhart told me
that he and Michael Turner were getting $\delta \rho/\rho \sim
10^{-16}$.  The second line refers to Stephen Hawking's seminar
in Princeton on June 7, when he announced the result that $\delta
\rho / \rho \sim 10^{-4}$.  The following line shows my own
answer, $\delta \rho/\rho \sim 50$, listed as a private
communication from me to me, on June 21.  Now I'll admit that
showing this transparency may have been a bit egotistical on my
part, but after all, these were the days when Stephen and I were
both young.

I was working with So-Young Pi, and as I said we started from
Hawking's work, struggling to fill in all the details that we
needed to get from one line of Stephen's preprint to the next. 
But we evaluated the time-derivative of the scalar field
differently from Stephen, and concluded that in the minimal SU(5)
model the perturbations would be far too large, with $\delta \rho
/ \rho \sim 50$ .  So-Young and I had started these calculations
together, but we didn't have the final answer until I completed
the calculations the day after I arrived here in Cambridge.
Cambridge is a great place to do calculations.  They gave us
rooms in Sidney Sussex College with thick walls and thick doors
and no telephones, so I worked late into the night and finished
the calculation the next morning, after breakfast, behind those
thick doors.  I have always thought Cambridge is a fantastic
place to do physics.  Anyway I got the answer shown in the third
line of the chart, $\delta \rho/\rho \sim 50$, which means that
the model did not work.  It is interesting to wonder whether I
was disappointed with my result, since inflation was my baby. 
The answer is that, since inflation had not really caught on yet
in 1982, I was absolutely thrilled to find an answer that I was
pretty sure corrected a calculation of Stephen's, which to me was
more of a victory than if inflation had worked!  In the end I
have had some of the benefits of both options, since the minimal
SU(5) grand unified theory is now ruled out by the absence of
observed proton decays, and we can construct other particle
theory models that allow the density perturbation amplitude to
turn out right.

Continuing, the next line shows the first seminar on density
fluctuations at the Nuffield Workshop, by Alexei Starobinski, on
Monday June 28, at the beginning of the second week.  My chart
shows him at $\delta \rho / \rho \sim 10^{-2}$, but he later
clarified that he found that the perturbations were of order
unity, and hence the real conclusion is simply that perturbation
theory could not be trusted.

Then Stephen gave his seminar, on Tuesday June 29, a little more
than a week into the conference.  Before this time I had had one
conversation with Stephen about density perturbations, but that
conversation was cut off by the start of the next seminar; I
remember that this was my seminar so I did not think I should
miss it!  Stephen was continuing to argue that his original
preprint was correct, but Paul Steinhardt and I were having
trouble understanding Stephen's line of reasoning.  Stephen's
seminar was impishly titled ``The End of Inflation,'' which of
course could have meant that he was going to discuss how the
inflationary era would end, or it could have meant that he was
going to argue that the inflationary theory is dead.  Stephen
always liked to generate surprises.  When Stephen began his
seminar, Paul and I were prepared to pounce at the point where we
thought he had made a mistake, hoping that the ensuing discussion
could settle the issue.  But when Stephen reached this point in
the seminar, very near the end, he jolted us by not using the
argument we expected, but instead substituted a new calculation
which in fact agreed with our calculation for the time derivative
of the scalar field.  He then agreed with our answer, that the
density perturbations were indeed far too large.  He expressed
his answer as $\alpha/C$, where $\alpha$ is the fine-structure
constant of the grand unified theory, and $C$ was a dimensionless
constant that was not calculated, which he referred to in his
talk as a ``fudge factor.''  In his talk, Stephen concluded that
we would need $C \sim 100$, or else the inflationary scenario
would have to be abandoned.  In the published version of
Stephen's paper \cite{Hawking-pub}, the abstract began
identically but ended differently:
\begin{quote} 
These [irregularities in the scalar field] would lead to
fluctuations in the rate of expansion which would have the right
spectrum to account for the existence of galaxies.  However the
amplitude would be too high to be consistent with observations of
the isotropy of the microwave background unless the effective
coupling constant of the Higgs scalar was very small. 
\end{quote} 
`Very small', as we now know, means about $10^{-13}$ or something
close to that.  I have never known when it was that Stephen
changed his mind, or whether he circulated this preprint just to
tease us from the beginning.  He certainly did not change his
mind while he was giving his talk, so he must have decided
earlier that his original preprint had the wrong answer.  But he
did not tell any of us.  Stephen recently told me that he
initially calculated the time derivative of the scalar field when
it was halfway down the hill, but talking to people at the
workshop made him realize that it should be evaluated at first
Hubble crossing, which gives much larger perturbations.  When
Stephen gave the talk, he proceeded without a pause, with no
suggestion that the result he was presenting was in contradiction
of his own preprint.

As my chart shows, by the time of my talk Bardeen, Steinhardt,
and Turner had invented a more accurate way to solve their
equations, and found an answer in agreement with the rest of us. 
The result that we all found was roughly
\be
\left( {\delta \rho \over \rho} \right)_{2t=\ell} = C \, {H \,
\delta \phi \over \dot \phi_{\rm cl}} \ ,
\label{delta-rho}
\ee
where $\delta \phi \sim H$ is the quantum fluctuation in $\phi$,
$C$ is a constant of order unity which depends on detailed
normalization conventions, and $\dot \phi_{\rm cl}$ is the proper
time-derivative of the function that describes the classical
evolution of the inflaton field.  The right-hand side is
evaluated at first Hubble crossing, the instant during the
inflationary era when the physical wavelength of the mode under
consideration is equal to $H^{-1}$.

\section{Observational evidence for inflation}

At the time of the Nuffield meeting, I thought it was great fun
and excitement to try to calculate the predictions for the
density perturbations resulting from inflation, but I never
believed that anyone would actually measure these density
perturbations.  I guess I am speaking mainly for myself, but I
suspect that the others felt the same way.  I knew that we had
some idea of what density perturbations were needed to explain
galaxy formation, and I'm sure I believed that such estimates
would improve with time.  But I never thought we would have
direct measurements of the density perturbations coming from the
microwave background.  The radiation itself is incredibly weak;
it is, we should keep in mind, thermal radiation at 3$\,$K, which
is about 100,000,000 times weaker than thermal radiation at room
temperature, which is itself pretty weak.  And the
nonuniformities that we are talking about are at the level of
only about one part in 100,000.  So I found it absolutely
astounding when the COBE results were published in 1992, and I
was even more astounded by the recent results of Boomerang,
Maxima, DASI, and CBI. 

\begin{figure}[p]
\centerline{\epsfbox{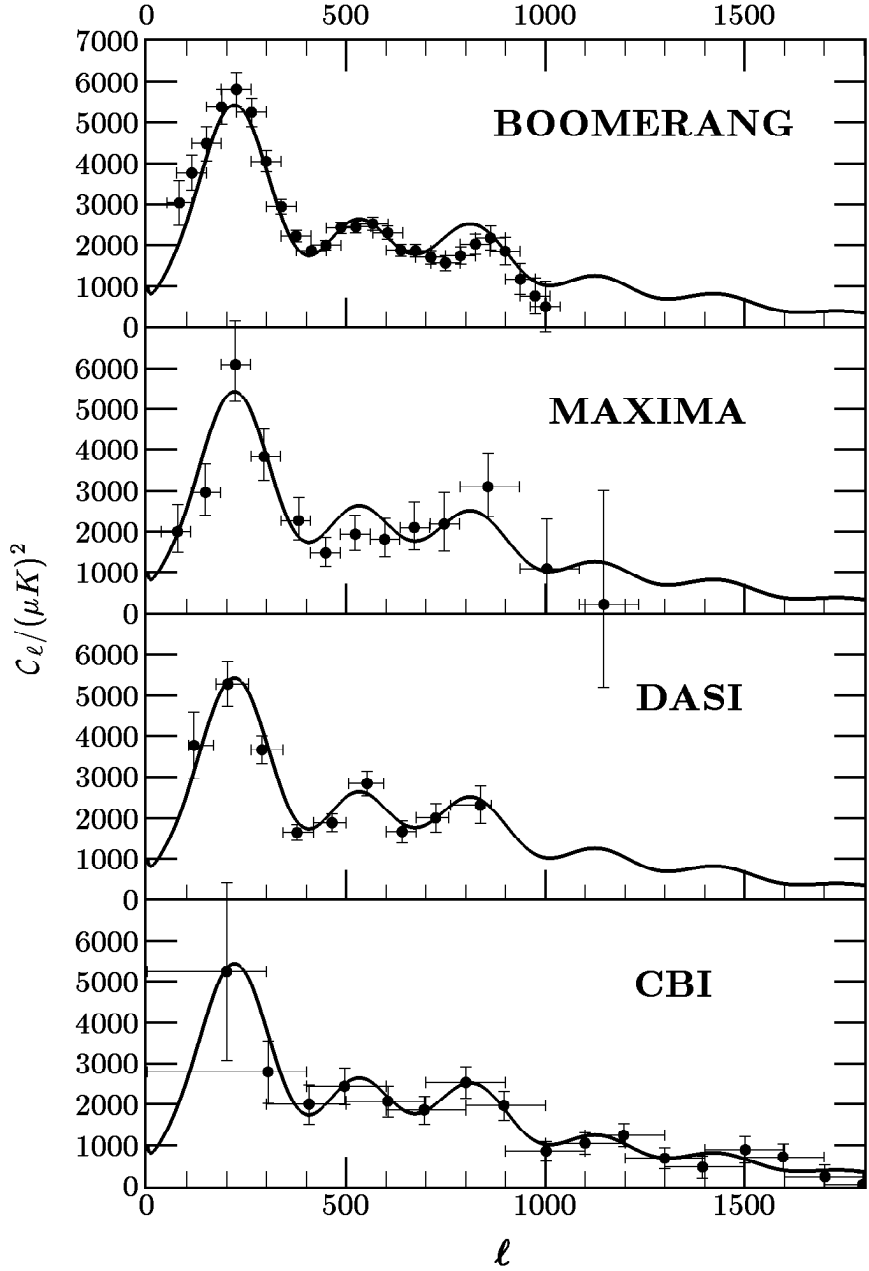}}
\caption{Angular power spectrum of the CMB compared to an
inflationary model.  The theoretical model is the best fit to all
the data obtained by the CBI group (\protect\cite{CBI}),
described by the parameters $\Omega_{\rm tot}=1.0$,
$\Omega_{\Lambda}=0.7$, $\Omega_{\rm CDM}=0.257$, $\Omega_b h^2 =
0.020$, $h=0.68$, $n_s=0.95$, $\tau_c=0$.} 
\figurelabel{CMBdata}
\end{figure}

Today the measurements of the cosmic microwave background (CMB)
and the perturbations that they exhibit are truly extraordinary. 
\figuremark{CMBdata} is a graph of the most recent data that was
released by the Boomerang group \cite{BOOMERANG}, the Maxima
group \cite{MAXIMA}, the DASI group \cite{DASI}, and the CBI
group \cite{CBI}.  These experiments measure the nonuniformities
$\Delta T(\theta,\phi)$ of the CMB radiation, which are described
by their expansion in spherical harmonics:
\be
\Delta T(\theta,\phi) = \sum_{\ell m} a_{\ell m} Y_{\ell m} (
\theta, \phi ) .
\ee
For each $\ell$ one defines
\be
C_\ell \equiv \left\langle \left\vert a_{\ell m} \right\vert^2
\right\rangle ,
\ee
and then
\be
{\cal C}_\ell \equiv {1 \over 2 \pi} \ell (\ell + 1) C_\ell .
\ee
The graphs show the measurements of the ${\cal C}_\ell$'s in
microkelvin$^2$, as a function of multipole $\ell$.  The graph
also shows a theoretical prediction, corresponding to the ``Joint
model'' of Ref.~\cite{CBI}, which was computed using the computer
code of Seljak and Zaldarriaga, CMBFAST \cite{CMBFAST}.

The curves show a series of well-defined ``acoustic'' peaks,
caused by the oscillations of the cosmic fluid of photons,
neutrinos, electrons, protons, and dark matter.  Roughly speaking
the image of the cosmic background radiation on the sky is a
snapshot of what the universe looked like at the time of last
scattering, a time about 400,000 years after the big bang, at
which the plasma of the early universe neutralized to form a
transparent gas.  Although the fluctuations of all wavelengths
are imprinted on the CMB at the same time, it is nonetheless a
conceptually useful approximation to think of the graph of ${\cal
C}_\ell$'s versus $\ell$ as if it were a graph of the
fluctuations of the early universe as a function of time.  The
shorter wavelength (i.e., higher $\ell$) fluctuations evolve
faster, so at the time of last scattering they have undergone
more oscillations and are therefore in a later stage of their
evolution than their longer wavelength cousins.  The lower
amplitude at higher $\ell$ is partly due to the damping of these
oscillations, and partly due to the fact that the time of last
scattering is not unique, since the transition from opacity to
transparency is actually somewhat gradual.  Thus the CMB image
has some similarities to a photographic time exposure, and the
small scale details are therefore blurred.

\begin{figure}[htb]
\centerline{\epsfbox{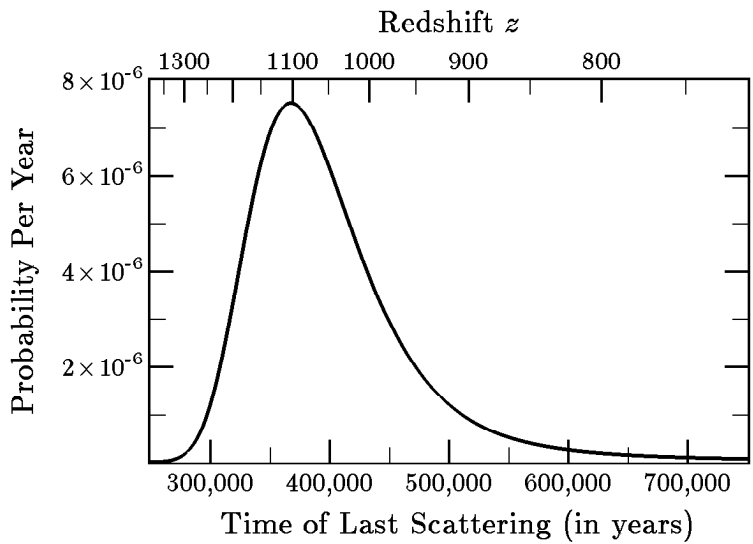}}
\caption{The probability distribution for the time of last
scattering of the CMB photons.  The curve was computed using the
same parameters as in \figuremark{CMBdata}.} 
\figurelabel{LastScattering}
\end{figure}

The probability distribution for the time of last scattering is
shown in \figuremark{LastScattering}, calculated\footnote{The
visibility function shown here was calculated using CMBFAST, but
the program had to be modified to print out this information,
which normally is used internally.  I also modified the code
slightly to keep track of proper time, as opposed to conformal
time.  The ionization evolution was calculated using the RECFAST
subroutine distributed with CMBFAST, which was written by
D.~Scott, based on calculations by Seager, Sasselov, and Scott
\cite{RECFAST}.} for exactly the same parameters as the curves in
\figuremark{CMBdata}.  The median time of last scattering is
388,000 years, the peak of the curve is at 367,000 years, and the
full width at half maximum is 113,000 years.  (The mean was
computed as 475,000 years, but the calculation was influenced
significantly by a long late-time tail of the probability
distribution, extending to billions of years; it is not clear if
this calculation is reliable or relevant.)

The theoretical curve shown in \figuremark{CMBdata} has an
amplitude which is normalized to the data; in practice
inflationary models do not make any prediction for the overall
amplitude of the fluctuations, though we could in principle make
a prediction if we really knew the potential energy function of
the inflaton field, the scalar field that drives inflation. 

However, everything else about this curve is pretty well fixed
either by inflation, or by astronomical determinations of
cosmological parameters.  For example, the curves depend
sensitively on $\Omega_{\rm tot} = \rho_{\rm tot}/\rho_c$, where
$\rho_{\rm tot}$ is the total mass density, and the critical mass
density $\rho_c$ is defined to be that density which corresponds
to a flat spatial geometry.\footnote{Until recently it was common
to say that the critical density was that density which put the
universe just on the borderline between eternal expansion and
eventual collapse.  This definition was never generally accepted
as a technical definition, but was nonetheless often used in
lectures, especially those intended for a non-technical audience. 
If the only materials present are normal matter, dark matter, and
radiation, then the two definitions are equivalent.  However,
since we now believe that the universe contains a large amount of
``dark energy'' with negative pressure, these definitions are no
longer equivalent.  We must therefore keep in mind that the
critical density is defined in terms of the spatial geometry of
the universe.} $\rho_c$ is related to the Hubble constant $H$ by
$\rho_c = 3 H^2 / (8 \pi G)$, where $G$ is Newton's gravitational
constant.  Inflation predicts that $\Omega_{\rm tot}=1$, and the
curves are drawn for $\Omega_{\rm tot}=1$.  If $\Omega_{\rm tot}$
were decreased from 1, all the peaks would shift to the right. 
If $\Omega_{\rm tot}$ were 0.3, as was widely believed five years
ago, the first peak would be at about $\ell = 400$. 

The curve also depends on the present value of the Hubble
constant, which was taken to be $H \equiv 100\, h \cdot
$km$\cdot$sec$^{-1}\cdot$Mpc$^{-1}$, with $h=0.68$.  This value
is completely consistent with the Hubble Key Project value
\cite{Freedman}, $h = 0.72 \pm 0.08 $.  An increase in $h$ would
push all the peaks downward and toward the left.  

The density of baryons in the universe also affects the predicted
curves.  This density is usually quantified by the product
$\Omega_b h^2$.  (Note that $\Omega_b h^2 \equiv \rho_b h^2 /
\rho_c$, so the explicit factor of $h^2$ cancels the
$h$-dependence of $\rho_c$, and consequently $\Omega_b h^2$ is
really just an indirect way of describing $\rho_b$.)  A high
density of baryons suppresses the second peak, and indeed when
the first Boomerang results \cite{BOOMERANG1} were released in
2000, the absence of an apparent second peak led some to
speculate (see, for example, Ref.~\cite{Hu2000}) that the
universe might contain 50\% more baryons than expected on the
basis of big bang nucleosynthesis.  This problem soon dissolved
with the appearance of new data, and the fit shown in
\figuremark{CMBdata} uses $\Omega_b h^2 = 0.020$, in perfect
agreement with nucleosynthesis-based estimate of Burles, Nollett,
and Turner \cite{BNT}, $\Omega_b h^2 = 0.020
\pm 0.002$ (95\% confidence).  

The scalar power law index $n_s$ is predicted to be very near one
for simple inflationary models, corresponding to a
scale-invariant Harrison-Zeldovich \cite{Harrison,Zeldovich}
spectrum; the data is fit with $n_s = 0.95$.  $\tau_c=0$ means
that no re-ionization of the intergalactic medium is assumed, and
the values used for the densities of cold dark matter
($\Omega_{\rm CDM} = 0.257$) and ``dark energy''
($\Omega_\Lambda=0.7$) are consistent with the supernova
observations \cite{High-Z, SupernovaCosProj}.

So, except for the height of the curve, the data shown in
\figuremark{CMBdata} can be essentially predicted on the basis of
other measurements and on the inflationary model.  I consider
this a spectacular success.  I am amazed that this data can be
measured, and that it agrees so well with what we predicted back
in 1982.  It seems to show not only that inflation is correct,
but that the simplest form of slow-roll inflation is correct. 
Given the amount of flexibility that inflationary models allow,
and also the uncertainties that exist in the data, I suspect that
probably it is fortuitous that the agreement is as good as what
we see.  My guess is that we are going to see some disagreements
before things really fit together tightly, but nonetheless we are
currently seeing a very spectacular agreement which strongly
suggests that we are on the right course. 

\section{Eternal inflation}
For the remaining half of my talk, I want to discuss an issue
related to the eternal nature of inflation. This may be only
vaguely related to the topic that I was asked to talk about,
inflation and cosmological perturbations, but there is a
connection.  The density perturbations arising from inflation are
never truly scale invariant, but typically grow slowly at large
wavelengths.  Eternal inflation is really a consequence of the
very long wavelength, high amplitude tail of the density
perturbation spectrum of inflation.  At extremely long
wavelengths these perturbations are usually so large that they
prevent inflation from ending at all in some places, leading to
what is called eternal inflation. 

My main goal is to describe a singularity theorem that I recently
proved in collaboration with Alex Vilenkin and Arvind Borde
\cite{BGV}.  But first, I would like to give a brief explanation
of how eternal inflation works.

There are basically two versions of inflation, and consequently
two answers to the question of why inflation is eternal (see
\figuremark{inftypes}).  In the case of new inflation, the
exponential expansion occurs as the scalar field rolls from the
peak of a potential energy diagram down to a trough.  (I refer to
the state for which the scalar field is at the top of the hill as
a {\em false vacuum}, although this is a slight change from the
original definition of the phrase.)  The eternal aspect occurs
while the scalar field is hovering around the peak.  The first
model of this type was constructed by Steinhardt
\cite{Steinhardt-eternal} in 1983, and later that year Vilenkin
\cite{Vilenkin-eternal} showed that new inflationary models are
generically eternal.  The key point is that, even though
classically the field would roll off the hill,
quantum-mechanically there is always an amplitude, a tail of the
wave function, for it to remain at the top.  If you ask how fast
does this tail of the wave function fall off with time, the
answer in almost any model is that it falls off exponentially
with time, just like the decay of most metastable states
\cite{Guth-Pi2}.  The time scale for the decay of the false
vacuum is controlled by
\be
m^2 = -\left.{\partial^2 V \over \partial \phi^2}\right|_{\phi=0} \ ,
\ee
the negative mass-squared of the scalar field when it is at the
top of the hill in the potential diagram.  This is an adjustable
parameter as far as our use of the model is concerned, but $m$
has to be small compared to the Hubble constant or else the model
does not lead to enough inflation.  So, for parameters that are
chosen to make the inflationary model work, the exponential decay
of the false vacuum is slower than the exponential expansion. 
Even though the false vacuum is decaying, the expansion outruns
the decay and the total volume of false vacuum actually increases
with time rather than decreases.  Thus inflation does not end
everywhere at once, but instead inflation ends in localized
patches, in a succession that continues ad infinitum.  Each patch
is essentially a whole universe --- at least its residents will
consider it a whole universe --- and so inflation can be said to
produce not just one universe, but an infinite number of
universes.  These universes are sometimes called bubble
universes, but I prefer to use the phrase ``pocket universe,'' to
avoid the implication that they are approximately round.  (While
bubbles formed in first-order phase transitions are round
\cite{coleman-deluccia}, the local universes formed in eternal
new inflation are generally very irregular, as can be seen for
example in the two-dimensional simulation by Vanchurin, Vilenkin,
and Winitzki in Fig.~2 of Ref.~\cite{vvw}.)

\begin{figure}[ht]
\centerline{\epsfbox{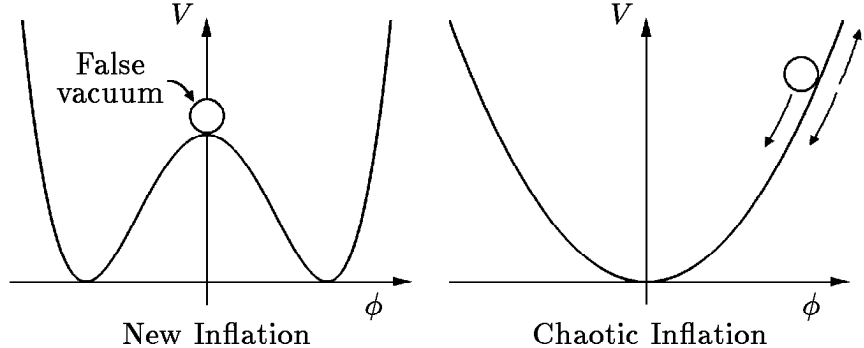}}
\caption{Eternal inflation from both a new inflationary potential
(left) and a chaotic inflationary potential (right).}
\figurelabel{inftypes}
\end{figure}

In the context of chaotic inflationary models, as developed by
Andrei Linde (who both proposed the models \cite{Linde-chaotic1,
Linde-chaotic2} and showed that they are eternal
\cite{Linde-eternal1, Linde-eternal2, Goncharov, Linde-book}),
the situation is slightly more complicated. Inflation is
occurring as the scalar field rolls down a hill of the potential
energy diagram, starting high on the hill.  As the field rolls
down the hill, quantum fluctuations will be superimposed on top
of the classical motion.  The best way to think about this is to
ask what happens during one time interval of duration $\Delta t =
H^{-1}$ (one Hubble time), in a region of one Hubble volume
$H^{-3}$.  Suppose that $\phi_0$ is the average value of $\phi$
in this region, at the start of the time interval.  By the
definition of a Hubble time, we know how much expansion is going
to occur during the time interval: exactly a factor of $e$. 
(This is the only exact number in today's talk, so I wanted to
emphasize the point.)  That means the volume will expand by a
factor of $e^3$.  One of the deep truths that one learns by
working on inflation is that $e^3$ is about equal to 20, so the
volume will expand by a factor of 20.  Since correlations
typically extend over about a Hubble length, by the end of one
Hubble time, the initial Hubble-sized region grows and breaks up
into 20 independent Hubble-sized regions. 

As the scalar field is classically rolling down the hill, the
classical change in the field $\Delta \phi_{\rm cl}$ during the
time interval $\Delta t$ is going to be modified by quantum
fluctuations $\Delta \phi_{\rm qu}$, which can drive the field
upward or downward relative to the classical trajectory.  For any
one of the 20 regions at the end of the time interval, we can
describe the change in $\phi$ during the interval by
\be
\Delta \phi = \Delta \phi_{\rm cl} + \Delta \phi_{\rm qu} \ . 
\ee
In lowest order perturbation theory the fluctuation is treated as
a free quantum field, which implies that $\Delta \phi_{\rm qu}$,
the quantum fluctuation averaged over one of the 20 Hubble
volumes at the end, will have a Gaussian probability
distribution, with a width of order $H/2 \pi$
\cite{Vilenkin-Ford, Linde-random, Starobinsky,
Starobinsky-random}.  There is then always some probability that
the sum of the two terms on the right-hand side will be positive
--- that the scalar field will fluctuate up and not down.  As
long as that probability is bigger than 1 in 20, then the number
of inflating regions with $\phi \ge \phi_0$ will be larger at the
end of the time interval $\Delta t$ than it was at the beginning. 
This process will then go on forever, so inflation will never
end. 

Thus, the criterion for eternal inflation is that the probability
for the scalar field to go up must be bigger than $1/e^3 \approx
1/20$.  For a Gaussian probability distribution, this condition
will be met provided that the standard deviation for $\Delta
\phi_{\rm qu} $ is bigger than $0.61 |\Delta \phi_{\rm cl}|$.
Using $\Delta \phi_{\rm cl} \approx \dot \phi_{\rm cl} H^{-1}$,
the criterion becomes
\be
\Delta \phi_{\rm qu} \approx {H \over 2 \pi} > 0.61 \, | \dot
\phi_{\rm cl}| \, H^{-1} \Longleftrightarrow {H^2 \over |\dot
\phi_{\rm cl}|} > 3.8 \ .
\label{Criterion}
\ee
Comparing with Eq.~(\ref{delta-rho}), we see that the condition
for eternal inflation is equivalent to the condition that $\delta
\rho/\rho$ on ultra-long length scales is bigger than a number of
order unity.

The probability that $\Delta \phi$ is positive tends to increase
as one considers larger and larger values of $\phi$, so sooner or
later one reaches the point at which inflation becomes eternal. 
If one takes, for example, a scalar field with a potential
\be
V(\phi) = {1 \over 4} \lambda \phi^4 \ ,
\ee
then the de Sitter space equation of motion in flat
Robertson-Walker coordinates (${\rm d}s^2 = -{\rm d}t^2 + e^{2 H
t} {\rm d} \vec x^2$) takes the form
\be
\ddot \phi + 3 H \dot \phi = - \lambda \phi^3 \ ,
\ee
where spatial derivatives have been neglected.  In the
``slow-roll'' approximation one also neglects the $\ddot \phi$
term, so $\dot \phi \approx - \lambda \phi^3 / (3 H)$, where the
Hubble constant $H$ is related to the energy density by
\be
H^2 = {8 \pi \over 3} G \rho = {2 \pi \over 3} {\lambda \phi^4
\over M_p^2} \ ,
\ee
where $M_p \equiv 1/\sqrt{G}$ is the Planck mass.  Putting these
relations together, one finds that the criterion for eternal
inflation, Eq.~(\ref{Criterion}), becomes
\be
\phi > 0.75 \, \lambda^{-1/6} \, M_p \ .
\ee

Since $\lambda$ must be taken very small, on the order of
$10^{-12}$, for the density perturbations to have the right
magnitude, this value for the field is generally well above the
Planck scale.  The corresponding energy density, however, is
given by
\be
V(\phi) = {1 \over 4} \lambda \phi^4 = .079 \lambda^{1/3} M_p^4
\ ,
\ee
which is actually far below the Planck scale.

So for these reasons we think inflation is almost always eternal. 
I think the inevitability of eternal inflation in the context of
new inflation is really unassailable --- I do not see how you
could possibly avoid it, assuming that the rolling of the scalar
field off the top of the hill is slow enough to allow inflation
to be successful.  The argument in the case of chaotic inflation
is a bit more approximate, and some people have questioned it,
but I still believe it has to work because the criterion that has
to be satisfied is so mild.  For eternal inflation to set in, all
one needs is that the probability for the field to increase in a
given Hubble-sized volume during a Hubble time interval is larger
than 1/20.

\section{A new singularity theorem}
Eternal inflation implies that once inflation starts, it never
stops.  This leads to the question: can inflation by itself be
the complete theory of cosmic origins?  Can inflation be eternal
into the past as well as the future, allowing a model which on
very large scales is steady state, eliminating the need for a
beginning?  The answer I believe is no, although I would not
claim that we have a rock-solid proof.  Borde, Vilenkin, and I
\cite{BGV} have proven a rigorous theorem, which I will describe,
and this theorem certainly shows that the simplest type of
inflationary models still require a beginning, even though they
are eternal into the future.  The difficulty is that we have no
way of discussing the class of all possible inflationary models,
so we cannot say that our theorem applies to all cases.  Our
singularity theorem was certainly inspired by Stephen's famous
work on singularity theorems, which established the value of such
theorems, so it is very fitting that I describe this theorem at a
symposium in honor of Stephen's 60th birthday.  And I am sure
that our conclusions fall on the side that Stephen would prefer,
since Stephen has put much effort into studying the quantum
origin of the universe, a subject which could have been bypassed
if inflationary models could avoid a beginning.

I will not try to state the theorem immediately, since it will be
easier later, after some definitions have been put forward.

An unusual feature of the new singularity theorem is that it
avoids any mention of things such as energy conditions, which are
crucial assumptions for other formulations of singularity
theorems.  In particular, in the context of eternal inflation,
the key papers by Borde and Vilenkin \cite{Borde-Vilenkin1,
Borde, Borde-Vilenkin2, Borde-Vilenkin3} proved several different
theorems, all of which invoked the weak energy condition.  This
is the condition that $n_\mu n_\nu T^{\mu\nu} \ge 0$, where
$T^{\mu\nu}$ is the energy-momentum tensor and $n_\mu$ is any
timelike vector.  For a perfect fluid, the condition is
equivalent to assuming that $\rho \ge 0$ and $\rho + p \ge 0$,
where $\rho$ is the energy density and $p$ the pressure.  The
weak energy condition is always valid classically, but it is
nevertheless violated by quantum fluctuations.  In particular, in
a perfect de Sitter space the quantity $\rho + p$ is identically
equal to zero.  In a quantum description, however, the vacuum is
never an eigenstate of $\rho + p$, so the quantity must
fluctuate, but it must still average to its classical value of
zero.  It therefore fluctuates both positively and negatively,
thereby violating the weak energy condition about half the time
\cite{Borde-Vilenkin-fluctuations, GVW}.  The nice thing about
the new theorem is that it is purely kinematical; it really
depends in no way on the dynamics of general relativity, but only
on the redshifting of velocities in an expanding universe.  It
seems rather amazing, however, that we can learn something useful
by considering only relativistic kinematics.  Crudely speaking,
the theorem says that if the universe expands fast enough, then
it cannot possibly be geodesically complete to the past.  In a
few minutes I will be able to define what it means for the
universe to expand fast enough.

We wish to prove a theorem that will apply to any universe, no
matter how inhomogeneous or anisotropic, so we want to think of
the expansion as a local phenomenon.  To describe this expansion,
we need to adopt a local definition of the Hubble parameter.  One
way to define a local Hubble parameter would be to imagine
measuring the velocities of particles moving with the Hubble flow
within some small neighborhood, and then one could define a local
Hubble parameter in terms of the divergence of the velocity
field.  For the purpose of our theorem, however, we have found it
useful to consider an even more local definition of the Hubble
parameter, one that can be measured by a single geodesic observer
traveling through the universe.  Our hypothetical observer would
never get a job at the Keck because he does not know how to use a
telescope.  He is completely myopic, able to measure the velocity
only of those particles that intersect his own trajectory.  If he
is at rest relative to the local Hubble expansion he will not be
able to measure anything, but if he is moving relative to the
Hubble expansion he will pass a succession of particles, and will
infer a Hubble expansion parameter from the rate at which those
particles are separating from each other.  We will call the
particles that the observer passes ``comoving test particles,''
but for the purpose of the theorem it is not really necessary
that these particles exist.  All that is necessary is that the
worldlines of these hypothetical particles can be defined on the
background spacetime, and that each worldline has zero
acceleration at the instant that it intersects the observer's
trajectory.  The observer will be assumed to be traveling on a
geodesic, either timelike or null.  As illustrated in
\figuremark{definitions}, the four-velocity of the observer will
be called $v^\mu(\tau)$, where $\tau$ will denote proper time for
the case of a timelike observer, and an affine parameterization
for the case of a null observer.  In either case $v^\mu(\tau)
\equiv {\rm d} x^\mu / {\rm d} \tau $.  The comoving test
particle passed by the observer at time $\tau$ will be moving at
a four-velocity called $u^\mu(\tau)$.

\begin{figure}[ht]
\centerline{\epsfbox{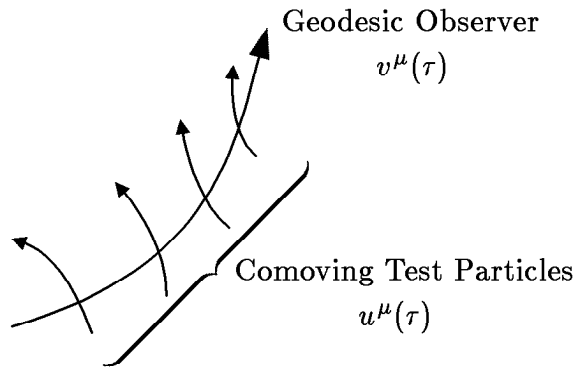}}
\caption{An observer measures the velocity of passing test
particles to infer the Hubble parameter.}
\figurelabel{definitions}
\end{figure}

To define the Hubble parameter that the observer measures at time
$\tau$, the observer focuses on two particles, one that he passes
at time $\tau$, and one at $\tau + \Delta \tau$, where in the end
he takes the limit $\Delta \tau \rightarrow 0$.  The Hubble
parameter is defined by
\be
H \equiv {\Delta v_{\rm radial} \over \Delta r} \ ,
\label{Hubble-def}
\ee
where $\Delta v_{\rm radial}$ is the radial component of the
relative velocity between the two particles, and $\Delta r$ is
their distance, where both quantities are computed in the rest
frame of one of the test particles, not in the rest frame of the
observer.  Note that this definition reduces to the usual one if
it is applied to a homogeneous isotropic universe, but it can be
defined for any spacetime in which one has identified a geodesic
observer and a family of comoving test particle trajectories.  We
will also be interested in the relative velocity between the test
particles and the observer, which can be measured by
\be
\gamma \equiv u_\mu v^\mu \ .
\ee
For the case of timelike observers, $\gamma$ corresponds to the
usual Lorentz factor of special relativity 
\be
\gamma = {1 \over \sqrt{1-v^2_{\rm rel}}} \ .
\ee

Now the key point is that the observer, if $H$ is positive, will
see himself redshifting in an expanding universe, which means
that his velocity relative to the test particles will be slowing
down.  Looking into the past, however, the velocity will become
blueshifted, becoming faster and faster as one follows the
trajectory further into the past.  We will find that under many
circumstances this velocity reaches the limiting speed of light
in a finite proper time, and then the trajectory can be continued
no further.

We are accustomed to calculating the redshifting of particle
velocities by writing the metric for the background spacetime and
then solving the geodesic equations.  However, the slowdown is
a kinematical effect that is much simpler than one might infer
from seeing the standard calculation.  To understand the logic,
it is useful to begin with a simple one-dimensional
nonrelativistic analogy.  Consider a stream of cars traveling on
a straight road, and imagine that they are moving apart from each
other, so the cars are the comoving test particles of an
expanding universe of cars.  You will play the role of the
geodesic observer in this thought experiment, and to simplify the
description we can imagine that the rest frame of the road
coincides with your rest frame; i.e., you are standing at rest
just alongside the road.  Suppose that one car goes by and you
measure its speed relative to you as 40 miles per hour.  When the
next car comes by, you know that its speed is {\em not} going to
be 50 mph, because that would mean it was catching up to the first
car, and we already said that the cars were getting further
apart.  If they are getting further apart, then the second car
has to be moving slower than the first, and the rate that they
are getting further apart is directly proportional to the
difference between the speeds.  So the Hubble expansion rate that
you would measure for the stream of cars is directly proportional
to the rate at which the relative speed between you and the cars
is decreasing.

When one increases the number of space dimensions from one to
three, one might expect that the simple relationship would break
down.  Now the velocity of the comoving test particles relative
to the observer can change both in magnitude and direction, so
the result is less obvious.  However, if one remembers that the
Hubble parameter is defined in terms of the radial component of
the relative velocity, a short calculation shows that the
one-dimensional result is still valid: the Hubble parameter is
directly proportional to the rate of change of the relative speed
between the observer and the comoving test particles, with
changes in direction giving no contribution.  Finally, one wants
to generalize the calculation to allow for relativistic
velocities.  As Borde, Vilenkin, and I showed in our paper, one
finds again that the Hubble parameter as defined by
Eq.~(\ref{Hubble-def}) can be related directly to the rate of
change of the relative velocities.  Specifically, in our paper we
defined a quantity
\be
F(\gamma) = \left\{ \begin{array} {ll} 
            \gamma^{-1} & \hbox{for null observers}\\
            {1 \over 2} \ln \left( {\gamma + 1 \over \gamma -1 }
            \right) & \hbox{for timelike observers,} 
            \end{array} \right.
\label{slowness}
\ee
which I like to call the slowness parameter.  Note that for the
timelike case, a relative velocity of zero ($\gamma=1$) corresponds
to infinite slowness ($F(\gamma)=\infty$), and that in both cases
the limiting relative velocity $\gamma \rightarrow \infty$
corresponds to zero slowness ($F(\gamma)=0$).  The existence of a
maximum possible relative velocity corresponds to a minimum possible
slowness.  The definition (\ref{slowness}) has been chosen so that
the minimum value is zero.

In our paper we showed that the Hubble parameter is related to
the slowness by the simple relation,
\be
H = {{\rm d} F (\gamma) \over {\rm d} \tau} \ .
\label{H-thm}
\ee
That is, the Hubble parameter is equal to the rate of change of
the slowness.  If $H > 0$ the universe expands, and the slowness
increases, which means that the velocity of the geodesic observer
relative to the comoving test particles is redshifted.  If one
looks backwards along the geodesic, however, one sees a
blueshift.  The velocity of the geodesic observer relative to the
comoving test particles increases as one looks backwards, and the
slowness decreases.  However, the slowness cannot fall below
zero.  Once the slowness reaches zero it cannot get any lower, so
according to Eq.~(\ref{H-thm}) it is impossible to continue the
geodesic any further if $H$ continues to be positive.

The verbal argument of the previous paragraph can be translated
into a rigorous inequality by integrating Eq.~(\ref{H-thm}) over
proper time, and using the fact that the value of $F(\gamma)$ at
the lower limit of integration is always nonnegative.  For
timelike geodesics, this leads to the result
\be
\int^{\tau_f} \, H \, {\rm d} \tau \le {1 \over 2} \ln \left(
{\gamma + 1 \over \gamma -1 } \right) \ ,
\label{limit}
\ee
where $\gamma$ refers to the value of $u_\mu v^\mu $ at the
final time $\tau_f$, and the inequality holds for any value of
the lower limit of integration.  Thus, the integral of $H$ along
a backwards-going geodesic is limited by an expression that
depends only on the final value of the relative velocity between
the geodesic observer and the comoving test particles.  The limit
disappears only for the case in which the geodesic observer is at
rest (relative to the comoving test particles) at the final time,
in which case the right-hand-side of Eq.~(\ref{limit}) is
infinite.  The right-hand-side of Eq.~(\ref{limit}) can be
algebraically rewritten as
\be
\int^{\tau_f} \, H \, {\rm d} \tau \le \ln \left( {1 \over v_{\rm
rel}} \right) + \ln \left( 1 + \gamma^{-1} \right) \ , 
\ee
where the first term dominates for low velocities, and the second
term dominates for large velocities.

For null observers the result looks even simpler, but it depends
on the normalization chosen for the affine parameter.  If we
normalize it by the convention that $\gamma = {\rm d} t / {\rm d}
\tau = 1$ at the final time $\tau_f$, where $t$ is the time
measured by comoving observers,  then the bound is simply
\be
\int^{\tau_f} \, H \, {\rm d} \tau \le 1 \ .
\label{limit-null}
\ee
Eqs.~(\ref{limit}) and (\ref{limit-null}) are technically our
final results, but I would like to say a few words about what we
think these results mean.

One illustration of the theorem is its application to de Sitter
space, especially as described in flat Robertson-Walker
coordinates: 
\be
{\rm d} s^2 = - {\rm d} t^2 + e^{2 \bar H t} {\rm d} \vec x^2 \ ,
\label{deSitter-flat}
\ee
where $\bar H$ is a constant.  Using the worldlines $\vec x =$
{\it constant} as comoving test particle trajectories, this
spacetime would give $H = \bar H$ at all points in the coordinate
system, independent of the geodesic observer worldine used to
define the measurement.  As is well-known, however, at least to
readers of Hawking and Ellis \cite{Hawking-Ellis}, these
coordinates cover only half of de Sitter space.  The locus
$t=-\infty$ describes a null hypersurface that forms the boundary
of the two halves.  The comoving geodesics ($\vec x = $ {\it
constant}) have infinite length within the half of de Sitter
space described by these coordinates, but our theorem implies (as
can be verified by direct calculation) that any other
backwards-going timelike geodesic reaches $t=- \infty$ in a
finite amount of proper time.  In fact, any noncomoving
backwards-going timelike geodesic reaches the $t=-\infty$
hypersurface with $\gamma=\infty$, and therefore just saturates
the bound implied by our theorem.  If we follow such a
backwards-going geodesic, it is of course possible to continue it
into the other half of the de Sitter space, since de Sitter space
as a whole is geodesically complete.  Such an extended geodesic
would have infinite proper length.  Furthermore, since de Sitter
space is homogeneous, at any location along the infinite
backwards-going geodesic it would be possible to construct a
neighborhood containing comoving test particle trajectories that
would make the universe appear to be expanding, with $H = \bar
H$.  Our theorem guarantees, however, that it is not possible to
define such comoving test particle trajectories globally, so that
$H=\bar H$ everywhere along the backwards-going geodesic. 
In fact, since the bound is saturated as the trajectory reaches
the $t=-\infty$ hypersurface, the theorem guarantees that $H$
must turn negative beyond this hypersurface, no matter how the
comoving test particle trajectories might be chosen.  Thus,
comoving test particle trajectories can be chosen so that $H =
\bar H$ anywhere in the de Sitter space, but this condition
cannot be enforced everywhere at once.

One simple description of the entire de Sitter space is the
closed universe description,
\be
{\rm d}s^2 = - {\rm d} t^2 + \cosh^2(\bar H t) \left\{ {{\rm d} r
^2 \over 1-r^2} + r^2 \left[{\rm d} \theta^2 + \sin^2 \theta \,
{\rm d} \phi^2\right] \right\} \ .
\ee
For the comoving test particle trajectories $(r, \theta, \phi) =
$ {\it constant}, $H = \bar H \tanh (\bar H t)$, and thus the
space is contracting at early times and expanding only at late
times.  This is geodesically complete, but does not constitute a
model of inflation that is eternal into the past, since in the
past the model is contracting and not inflating.  If the false
vacuum that supports the de Sitter space is only metastable, it
would decay completely during the infinite period of contraction,
so inflation would never take place.

To describe the implications of our theorem succinctly, it is
useful to define a concept that we call {\em uniformly bounded
expansion.}  A region of spacetime is said to be undergoing
uniformly bounded expansion if it is possible to define a
congruence of comoving test particle trajectories with the
property $H > H_{\rm min}$ everywhere in the region, for some
$H_{\rm min} > 0$.  The half of de Sitter space described by
Eq.~(\ref{deSitter-flat}) provides an example of a spacetime with
uniformly bounded expansion throughout.  The bounds of
Eqs.~(\ref{limit}) and (\ref{limit-null}) imply that any
spacetime that contains at least one noncomoving past-directed
geodesic exhibiting uniformly bounded expansion cannot be
geodesically complete in the past.

\section{The origin of the universe}

Now what does this say about the universe?  First let me point
out a few disclaimers, to make it clear what the theorem does not
say.  It does not imply that an eternally inflating model must
have a unique beginning, and it also does not imply that there is
an upper bound on the length of all backward-going geodesics from
a given point.  Our theorem places no bound on the length of the
comoving trajectories, and we know from the de Sitter space
example of Eq.~(\ref{deSitter-flat}) that such trajectories really
can be infinite.  Furthermore, our theorem allows for the
possibility of models that have regions of contraction
interspersed among regions of expansion, so that
Eqs.~(\ref{limit}) and (\ref{limit-null}) can perhaps be
satisfied for infinitely long backwards-going geodesics.  I
should also mention here for completeness that Aguirre and
Gratton \cite{Aguirre-Gratton} responded to our paper by
suggesting a geodesically complete model which uses the full de
Sitter space, but proposes that the thermodynamic arrow of time
is reversed in the two halves, so that both halves appear to be
expanding.  This successfully evades our theorem, but for my
taste it seems like an extravagant assumption.

In summary, the theorem by Borde, Vilenkin, and me does show that
any inflating model that is globally expanding must be
geodesically incomplete in the past.  This theorem certainly
applies to the simplest models of eternal inflation, in which an
approximately de Sitter region grows exponentially, with pieces
of it breaking off and decaying to form pocket universes. 
Geodesic incompleteness of the inflating region implies that
there must be some other physics introduced at the beginning, to
explain what happens at the past boundary of the inflating
region.  The most likely possibility, from my point of view, is
some kind of quantum origin, which of course touches on Stephen's
work.  This means that even in eternally inflating models, a
beginning is necessary.  I would expect that the details of this
beginning would be washed out by the inflationary evolution, but
nonetheless the model still requires a beginning of some sort,
something like the Hartle-Hawking \cite{Hartle-Hawking} or
Hawking-Turok \cite{Hawking-Turok} wavefunction of the universe.

%\section*{Acknowledgment}
\begin{acknowledgments}
The author was partially supported by the U.S. Department of
Energy (D.O.E.) under cooperative research agreement
\#DF-FC02-94ER40818.
\end{acknowledgments}

\end{document}